\documentclass[twocolumn]{aastex631}

\usepackage{graphicx}
\usepackage{booktabs}

\shorttitle{Variability of the {Na D}$_1$ Absorption Complex in $\eta$ Car}
\shortauthors{Pickett et al.}

\graphicspath{{./}{figures/}}

\newcommand{\kms}{km s$^{-1}$}

\begin{document}

\title{ Changes in the {Na D}$_1$ Absorption Components of $\eta$ Carinae Provide Clues on the Location of the Dissipating Central Occulter}

\correspondingauthor{Noel D. Richardson}
\email{noel.richardson@erau.edu}

\author[0000-0002-0684-4277]{Connor S. Pickett}
\affiliation{Department of Physics and Astronomy, Embry-Riddle Aeronautical University, 3700 Willow Creek Road, Prescott, AZ 86301, USA}

\author[0000-0002-2806-9339]{Noel D. Richardson}
\affiliation{Department of Physics and Astronomy, Embry-Riddle Aeronautical University, 3700 Willow Creek Road, Prescott, AZ 86301, USA}

\author[0000-0002-6851-5380]{Theodore Gull}
\affiliation{Exoplanets and Stellar Astrophysics Laboratory, NASA/Goddard Space Flight Center, Greenbelt, MD 20771}\affiliation{Space Telescope Science Institute, 3700 San Martin Drive. Baltimore, MD 21218, USA}

\author[0000-0001-5094-8017]{D. John Hillier}
\affiliation{Department of Physics and Astronomy and the Pittsburgh
Particle Physics, Astrophysics, and Cosmology Center
(PITT PACC), University of Pittsburgh, 3941 O’Hara Street, Pittsburgh,
PA 15260, USA}

\author[0000-0001-9853-2555]{Henrik Hartman}
\affiliation{Department of Materials Science and Applied Mathematics, Faculty of Technology and Society, Malm\"{o} University, SE-20506 Malm\"{o}, Sweden}

\author{Nour Ibrahim}
\affiliation{Department of Physics and Astronomy, Embry-Riddle Aeronautical University, 3700 Willow Creek Road, Prescott, AZ 86301, USA}
\affiliation{Department of Astronomy, University of Michigan, 1085 S. University, Ann Arbor, MI 48109, USA}

\author[0000-0003-0626-2717]{Alexis M. Lane}
\affiliation{Department of Physics and Astronomy, Embry-Riddle Aeronautical University, 3700 Willow Creek Road, Prescott, AZ 86301, USA}

\author{Emily Strawn}
\affiliation{Department of Physics and Astronomy, Embry-Riddle Aeronautical University, 3700 Willow Creek Road, Prescott, AZ 86301, USA}

\author[0000-0002-7978-2994]{Augusto Damineli}
\affiliation{Universidade de S\~ao Paulo, Instituto de Astronomia,
Geof\'isica e Ci\^encias Atmosf\'ericas, Rua do Mat\~ao 1226, Cidade Universit\'aria, S\~ao Paulo, Brasil}

\author[0000-0002-4333-9755]{Anthony F. J. Moffat}
\affiliation{D\'epartement de Physique and Centre de Recherche en Astrophysique du Qu\'ebec (CRAQ)
Universit\'e de Montr\'eal, C.P. 6128, Succ. Centre-Ville,
Montr\'eal, Qu\'ebec, H3C 3J7, Canada}

\author[0000-0002-0284-0578]{Felipe Navarete}
\affiliation{SOAR Telescope/NSF’s NOIRLab, Avda Juan Cisternas
1500, 1700000, La Serena, Chile}

\author[0000-0001-9754-2233]{Gerd Weigelt}
\affiliation{Max Planck Institute for Radio Astronomy, Auf dem
H\"ugel 69, D-53121 Bonn, Germany}

\begin{abstract}

The {Na D} absorption doublet in the spectrum of $\eta$ Carinae is complex, with multiple absorption features associated with the Great Eruption (1840s), the Lesser Eruption (1890s), and interstellar clouds. The velocity profile is further complicated by the P Cygni profile originating in the system's stellar winds and blending with the He I $\lambda$5876 profile. The Na D profile contains a multitude of absorption components, including those at velocities of $-$145 \kms, $-$168 \kms, and $+$87 \kms\ that we concentrate on in this analysis. Ground-based spectra recorded from 2008 to 2021 show significant variability of the $-$145 \kms\ absorption throughout long-term observations. In the high ionization phases of $\eta$ Carinae prior to the 2020 periastron passage, this feature disappeared completely but briefly reappeared across the 2020 periastron, along with a second absorption at $-$168 \kms. Over the past few decades, $\eta$ Car has been gradually brightening demonstrated to be caused by a dissipating  occulter. The decreasing absorption of the $-$145 \kms\ component, coupled with similar trends seen in absorptions of ultraviolet resonant lines, indicate that this central occulter was possibly a large clump associated with the Little Homunculus or another clump between the Little Homunculus and the star. We also report on a foreground absorption component at $+$87 \kms. Comparison of {Na D} absorption in the spectra of nearby systems demonstrates that this red-shifted component likely originates in an extended foreground structure consistent with a previous {ultraviolet} spectral survey in the Carina Nebula.

\end{abstract}


\section{Introduction} \label{sec:intro}

Luminous Blue Variables (LBVs) are evolved massive stars that are near the Eddington Limit, and are variable on time scales of days to centuries. The variability of these stars is such that when the star is observed being hotter, it is fainter in the optical and brighter in the ultraviolet. The majority of active LBVs tend to fluctuate between spectral types of {late-O or early }B- and F-type supergiants, while maintaining a nearly constant bolometric luminosity, with only small changes \citep[see for example the work on AG Car;][]{2009ApJ...698.1698G, 2011ApJ...736...46G}. These stars have been observed to have stellar eruptions where they eject {up to} several solar masses of material, creating circumstellar ejecta around many of these objects. The population of LBVs was recently documented in \citet{2018RNAAS...2..121R} and their properties have been reviewed in the past by \citet{1994PASP..106.1025H} and \citet{2001A&A...366..508V}.

$\eta$ Carinae {($\eta$ Car; HD 93308; HR 4210; previously $\eta$ Argus)} is a massive binary star within the Carina Nebula and is sometimes considered a prototype of the LBVs. The system is relatively close to us with a distance of {approximately} only 2.3 kpc \citep{2006ApJ...644.1151S}. $\eta$ Car has undergone two observed eruptions, causing significant amounts of gas and other debris to obscure the system itself, although there may have been additional eruptions in the past \citep{2016MNRAS.463..845K}. This material has resulted in a bipolar shell of ejecta surrounding the system from the Great Eruption in the 1840s and from the second {Lesser Eruption} in the 1890s, known as the Homunculus and the Little Homunculus, respectively. The binary is in a long-period 5.54 year orbit \citep[$2022.7\pm 0.3$ d;][]{1996ApJ...460L..49D, 1997NewA....2..107D, 2016ApJ...819..131T} that is highly eccentric \citep[$e=0.9$; e.g.,][]{2000ApJ...528L.101D,2020MNRAS.494...17G, 2022MNRAS.509..367G}. The primary star provides some ionization to the surrounding gas, such as causing Fe II emission lines, while the secondary star further ionizes the Homunculus and Little Homunculus. During the system's periastron passage, the secondary star passes into the far side of the optically thick wind of the primary star, which causes the ionization of gas in our line of sight to drop to normal levels for the primary star. This variability, where the ionization of the gas temporarily drops, has been referred to as spectroscopic events in the literature. As noted in several studies \citep[e.g.,][]{1982A&A...111..375M, 1998A&AS..133..299D, 2005AJ....129..900D, 2005A&A...435..183J, 2006ApJS..163..173G, 2010AJ....139.1534R, 2015AJ....150..109R}, the change in ionization causes increased absorption of low-ionization species from the ejecta, especially at velocities associated with the Little Homunculus (near $-$145 \kms).

In recent years, the central source of the $\eta$ Carinae system has been observed to be brightening. Unlike most normal LBV activity such as the S Doradus cycles, the central source is brightening in both the optical and ultraviolet in tandem. There are two competing hypotheses related to this variability. The first is that the central star is quickly evolving as it settles from the Great Eruption \citep{2010ApJ...717L..22M, 2011ApJ...740...80M, 2015A&A...578A.122M, 2019A&A...630L...6M, 2018ApJ...858..109D, 2021RNAAS...5..197M}. This suggestion relies on wind lines, such as H$\alpha$ \citep{2010ApJ...717L..22M}, changing with a long-term evolution, which could be interpreted as changes in the mass-loss properties. The second hypothesis is that there was an clump in the line of sight that has been dissipating and/or moving out of our line of sight \citep{1992A&A...262..153H, 2001ApJ...553..837H,2019MNRAS.484.1325D, 2021MNRAS.505..963D}. This is supported by the large amount of added extinction that is needed in the models of the stellar spectra by both \citet{2001ApJ...553..837H, 2006ApJ...642.1098H} and \citet{ 2012MNRAS.423.1623G}. As this extinction was fairly grey, a change in the extinction should allow most wavelengths to change at similar rates.

We aimed to use the variability of the {Na D} absorption component associated with the Lesser Eruption to understand how it relates to the documented changes in the spectrum of $\eta$ Carinae. The ionization potential of neutral Na is $\approx5.14$\,eV which corresponds to a wavelength of
$\sim$2412\AA. Thus the ionization of neutral sodium is achieved by radiation in the
{near ultraviolet}. However the ground state photoionization cross-section of Na I has a Cooper minimum\footnote{
A Cooper minimum is a region where the photoionization cross-section approaches zero,
and is typically exhibited by alkali elements \citep{1962PhRv..128..681C,2011aas..book.....P}} near 2000\,\AA\footnote{The Na I
photoionization data were computed by K T Taylor for the Opacity project \citep{Sea87_OP} and were obtained from http://cdsweb.u-strasbg.fr/topbase/home.html}, and consequently  a large fraction of the ionizations will occur by radiation shortward of 1800\,\AA. Given the large {ultraviolet} flux of $\eta$ Car, we expect Na to mostly singly ionized. As a consequence the time variation of the Na I equivalent widths will be  primarily governed by the time scale for the flux variations, and thus the orbital time scale around periastron. Thus we would expect the Na equivalent width ($\propto$column density) variations, and the {ultraviolet} flux variations to track each other. In Section 2, we describe 12 years of observations of the system from the Cerro Tololo Inter-American Observatory (CTIO) 1.5 m telescope and its echelle spectrographs, along with some complementary observations from the {\it Hubble Space Telescope} and the Space Telescope Imaging Spectrograph ({\it HST}/STIS). Section 3 details the measurements we made of three absorption components in the {Na D} spectrum. Section 4 discusses these findings, and we conclude this study in Section 5. 

\section{Observations} \label{sec:style}

Our observations come from a time-series of high-resolution spectroscopy of  $\eta$ Car taken with the CTIO 1.5 m SMARTS {(Small and Moderate Aperture Research Telescope System)} telescope. All data were taken with a fiber centered on the object, with a projected angular size of 2.7\arcsec. The first observations were taken in 2008--2010 with the fiber-fed spectrograph having a resolving power of $R\sim40,000$, and were described by \citet{2010AJ....139.1534R, 2015AJ....150..109R}. Since the 2009 periastron passage, a time-series of higher signal-to-noise and higher resolution spectra were obtained with the SMARTS telescope and the new CHIRON spectrograph with modes having a resolving power of $80,000 - 90,000$ for these observations \citep{2013PASP..125.1336T, 2021AJ....162..176P}. For the CHIRON time-series, the data were described by \citet{2016MNRAS.461.2540R}, with more observations having the same characteristics being taken in the intervening time, especially around the 2020 periastron passage. In total, we analyze 418 spectra taken with the CTIO 1.5 m telescope in this study.

{Data were previously reduced for all data previously published using techniques described in \citet{2010AJ....139.1534R, 2015AJ....150..109R} for the fiber echelle. This was done with standard techniques using IRAF\footnote{IRAF was distributed by the National Optical Astronomy Observatories, which is
operated by the Association of Universities for Research in Astronomy, Inc.,
under contract to the National Science Foundation.}. For the CHIRON data, the data were processed with the standard pipeline described recently by \citet{2021AJ....162..176P}. The observations also included a standard observation of HR 4468 (B9.5Vn) or $\mu$ Col (O9.5V), which was used to empirically derive the blaze function and then remove this through division with custom scripts in either IDL or with python. This was done in orders without spectral lines, with the remaining
orders (primarily around H$\alpha$ or H$\beta$) were interpolated to match
adjacent orders. 
}

The {observations} of the most recent periastron passage began with CTIO/CHIRON data collected on 15 October 2019 and concluded with data collected on 18 March 2020 when the telescope was closed due to the COVID-19 pandemic.
These new observations did provide {higher signal-to-noise} {Na D} profiles than ever before, owing to the {ongoing increase in the visual} brightness of $\eta$ Car \citep[e.g.,][]{2019MNRAS.484.1325D, 2021MNRAS.505..963D}.

\begin{figure*}[t!]
    \centering
    \includegraphics[width=7in]{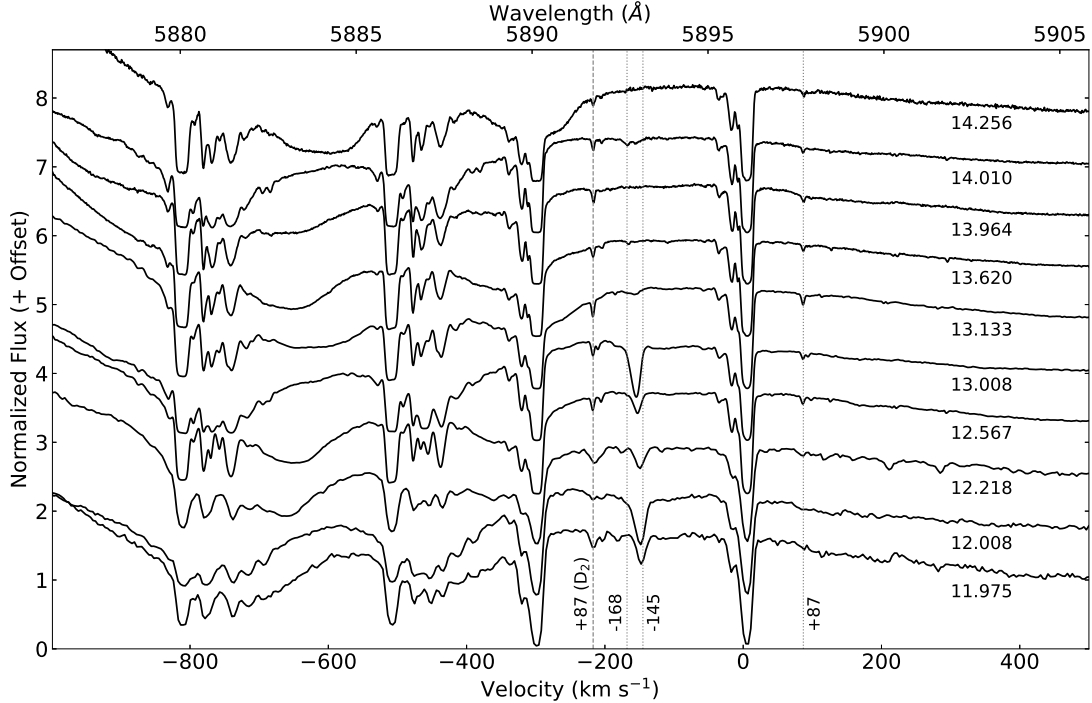}
    \caption{A subset of the data taken for $\eta$ Carinae, from phase $\phi = 11.975$ to $\phi = 14.256$ (indicated as numbers below the spectra on the right), which highlights some of the variations seen in the data. The dotted lines represent the multiple absorption features discussed in this paper relative to the Na D$_1$ line, with the $+87$ km s$^{-1}$ component also shown as a dashed line for the Na D$_2$ line. We indicate the velocity of the absorptions at the bottom of the figure. The data before phase 12.5 were taken with the lower-resolution fiber-fed echelle spectrograph, while the more recent data were taken with the CHIRON spectrograph. The variable emission on the red side of this plot is from \ion{He}{1} $\lambda$5876, for which the variability changes the apparent spacing on the left.}
    \label{fig 1: eta Car All}
\end{figure*}

In order to study the {Na D} complex, we first had to remove contamination from water in Earth's atmosphere. We utilized the telluric measurements from \citet{2003assi.book.....W} to construct a telluric template at the same resolution and sampling rate as the CHIRON data. This template could then be artificially strengthened or weakened to match each spectrum, under the assumption that the relative strengths of different lines remains fixed. This method was used over the entire {Na D} region, including regions with no intrinsic features from $\eta$ Carinae to ensure that the remaining spectrum was intrinsic to the source and the intervening material from the ejecta and the interstellar medium. 

When dealing with spectra of $\eta$ Car, it is important to use a reliable clock with which we can compare observations to the binary ephemeris. To that end, we adopt the ephemeris from \citet{2016ApJ...819..131T}, which should be close to the timing of periastron {in 2014}, and is given by the equation relating the phase $\phi$ to the heliocentric julian date (HJD) by

   $$ \phi = \frac{(HJD-2, 456, 874.4)}{2022.7} + 13.$$

In this context, the addition of the 13 in the equation means that the 2014 periastron was the thirteenth observed spectroscopic minimum since the first reported by \citet{1953ApJ...118..234G}.

We found an absorption component which we suspected to be interstellar in its origin, and we collected spectra of neighboring stars with the same high-resolution mode of the CHIRON spectrograph in July 2021. The objects observed were HD 93204, HD 93205, HD 303308, and CPD 59$^{\circ}$ 2628. Using the same methodology as with $\eta$ Car, the spectra were reduced and telluric corrected.

We also used archived spectra taken with the {\it Hubble Space Telescope}/Space Telescope Imaging Spectrograph ({\it HST}/STIS) in the E140M echelle mode with a nominal resolving power, R$=$45,800. The three spectra were recorded at phases $\phi =$ 11.119, 13.194 and 14.170, which is in the early recovery to the high-ionization state of the nebular structures. {These data were pipeline-processed and then downloaded from MAST.}

\section{Absorption Components at Various Velocities}

In Fig.~\ref{fig 1: eta Car All}, we show selected spectra from our time-series of the system. There are several features in this figure, which we will describe in detail in the following section. The interstellar components near 0 \kms\ (and near $-$300 \kms\ for the Na D$_2$ component in the shown Na D$_1$ reference frame) are both very close to black in our data, but some small uncorrectable amount of scattered light in the spectrographs makes the profile appear to have a minimum near 1\% of the normalized continuum light. 

There is a dramatic disappearance of a feature at $-$145 \kms\ {over our full data set}. {This and other velocities were fit for a radial velocity by means of a Gaussian profile or by measuring the point of minimum flux on select data sets with high signal-to-noise. We caution that these profiles are not always simple Gaussians, so we adopt these velocities to better describe the variations of the individual components.} \citet{2010AJ....139.1534R} described a similar component in the H$\alpha$ profile {that appeared and strengthened} across the 2009 periastron passage, and \citet{2015AJ....150..109R} noted changes in the {Na D} complex {similar to those of H$\alpha$ during the 2009 periastron passage}. In the most recent periastron passage, the feature's center has moved to bluer velocities, and the profile has bifurcated so that a second component at $-$168 \kms\ is now visible near periastron. Furthermore, there is a component at $+$87 \kms\ that is seen in all of the CHIRON spectra, without much variability present. For our analysis, we concentrate on the {Na D}$_1$ $\lambda 5896$ line, as the {Na D}$_2$ $\lambda 5890$ line is strongly blended with both the D$_1$ components originating in the Homunculus and the He I $\lambda 5876$ line.

\begin{figure} [t!]
    \centering
    \includegraphics[width=\columnwidth]{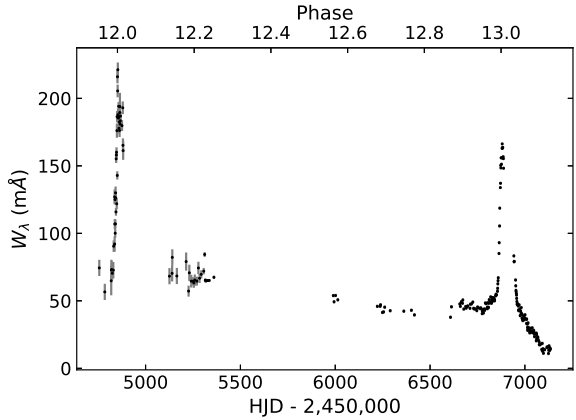}
    \caption{The measured equivalent width strength of the $-$145 \kms\ feature, in reference to Na D$_1$, through the 2009 and 2014 events, where the strength increases drastically near the two periastron passages. In the interim time, the feature is seen to have a slow decline, which accelerates to higher velocities after the 2014 event. }
    \label{fig 2: $-$145 \kms}
\end{figure}

For the features that we discuss in this paper, we measure the equivalent width defined as
$$    W_{\lambda} = \int{\frac{F_C(\lambda) - F_{\lambda}(\lambda)}{F_C(\lambda)}} d\lambda,$$
{where $F_C$ is the continuum flux and $F_\lambda$ is the flux as a function of wavelength ($\lambda$).}
The continuum around these features is difficult to define as the normalized spectrum shows broad {Na D} emission from the outer portions of the stellar wind. Therefore, we re-normalized each observation around the feature to measure the strength at that time of the observation. In order to estimate the errors in W$_{\lambda}$ of the measurements, we adopted the method of \citet{2006AN....327..862V}, where the error depends on the dispersion of the spectrum ($\Delta \lambda$), the number of spectral pixels integrated over ($n$), and the signal-to-noise ratio ($SNR$) of the spectrum per resolution element by the equation

$$\sigma = \frac{{n \Delta \lambda - W_\lambda}}{SNR}.$$ 

The errors for the weak feature at $-168$ \kms\ appear to be a bit larger than the scatter around the trends in the points even when we carefully measured the quantities needed {in the equation above}. Such a trend could mean that we over-estimated the errors, however since we had to locally normalize the data around the feature, within the emission component of the Na D complex, there are additional errors involved in that process. There are also sources of error that are not accounted for, such as how the strength could change on a changing slope of the emission profile. We have therefore kept these error bars at this level, since a measurement to a precision of $<5$m\AA\ is difficult to achieve with the star and this instrument. 

\subsection{The Absorption Component at -145 \kms} \label{subsec:tables}

\begin{figure} [t!]
    \centering
    \includegraphics[width=\columnwidth]{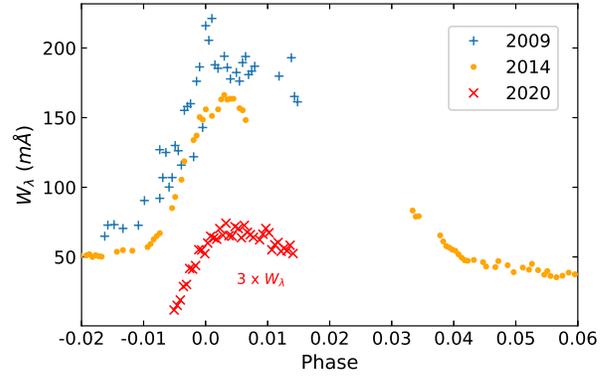}
    \caption{The measured equivalent width strength of the $-$145 \kms\ feature through the 2009, 2014, and 2020 events, phased to the binary clock. For the 2020 event, the equivalent width of the blended components at $-145$ and $-168$ \kms\ were measured, but multiplied by a factor of 3 to be displayed on a similar scale as the previous two periastron passages. }
    \label{fig 3: Zoomed-In Phases}
\end{figure}

\citet{2015AJ....150..109R} found that the absorption component near $-$145 \kms\ increases near the periastron passage, similar to the H$\alpha$ component that strengthens at the same time \citep{2010AJ....139.1534R, 2021MNRAS.505..963D}. We measured these spectra, along with the CTIO/CHIRON data, in the region surrounding the {variable} $-$145 \kms\ feature. The exact range changed with time as the feature moved to bluer velocities in more recent observations, {but the range of integration was typically about 25 \kms\ wide and moved with the absorption with time}. The equivalent width measurements are tabulated {in the online appendix} (Table \ref{table:145}). \citet{2006ApJS..163..173G} reported that ultraviolet features of this velocity seemed to appear stronger at periastron, yet remain fairly consistent in equivalent width ($W_{\lambda}$) throughout the remainder of the orbit. 

From our long time-series of measurements in Fig.~\ref{fig 2: $-$145 \kms}, the absorption at $-$145 \kms\ is observed to be weakening over this time period. In Fig.~\ref{fig 1: eta Car All} we see that this feature seems to move to bluer velocities over this same time period, while the interstellar absorption components stay constant in velocity as expected. By a time shortly after the 2014 periastron event, this feature was indiscernible in the spectra despite the higher signal-to-noise (S/N$\sim$150--200 per pixel). We see similarities in the two events for 2009 and 2014 when phased to the binary clock (Fig.~\ref{fig 3: Zoomed-In Phases}). The change in the ultraviolet flux during the periastron passage is about a factor of three according to \citet{2021ApJ...923..102G}, consistent with the changes seen in this line. The 2020 event is discussed in the following subsection.

It was reported in \cite{2006ApJS..163..173G} that the $-$145 \kms\ features in the {ultraviolet} would change intensity across the spectroscopic minimum. The absorption feature in Fe II, as measured at $\phi = 0.739$, appears during the broad maximum. However, when at a minimum, the feature appeared stronger and was also accompanied by two more features at velocities near $-$145 \kms. As noted by \citet{2006ApJS..163..173G}, this indicates a brief drop in ionization of the Little Homunculus in our line of sight caused by the shadowing of that region by the primary's wind intervening in front of the secondary star.

\subsection{The Absorption Component at -168 \kms}

\begin{figure}[t!]
    \centering
    \includegraphics[width=\columnwidth]{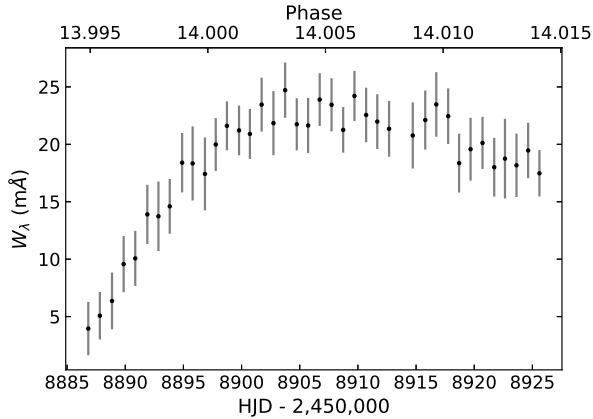}
    \caption{The measured equivalent width strength of the $-$168 \kms\ feature across the 2020 periastron passage, where the strength increases after periastron. This measurement is a blend of the bluer component of the $-$145 \kms\ with the $-$168 \kms\ component.}
    \label{fig 4: $-$168 \kms}
\end{figure}

During the recent 2020 periastron passage, an absorption component at $-$168 \kms\ has appeared. This absorption may have been present in earlier periastra, but would have been blended with a much stronger component at $-145$ \kms. It was also seen in the H$\alpha$ line as observed by \citet{2021MNRAS.505..963D}, although at the velocity of $-180$ \kms. {This change in the H$\alpha$ and Na D absorptions associated with the periastron passages is likely due to a changing ionization structure in the Little Homunculus.} A similar velocity component was reported in low-ionization ultraviolet lines by \citet{2006ApJS..163..173G} during the low-state corresponding to the periastron passages. 
This component, while independent of the $-$145 \kms\ feature, {seems to have become similar in strength to its changing-velocity counterpart at $-145$ \kms, and could have been undetected in previous cycles due to the stronger component at $-145$ \kms.} We compare the blended component to that of the $-145$ \kms\ component in Fig.~\ref{fig 3: Zoomed-In Phases}, where we see that the beginning of the increase in absorption shows a similar time from the pre-event minimum to maximum as the previous events.

While not as prominent as the $-$145 \kms\ feature had been, the $-$168 \kms\ feature became significantly strong enough to take equivalent width measurements during the last periastron. Measurements for the $-$168 \kms\ feature were taken between the wavelengths of 5892.45{\AA} and 5893.20{\AA}, corresponding to the radial velocities of $-$180.82 \kms\ and $-$142.62 \kms\, respectively. This integration includes the smaller contribution of the $-$145 \kms\ component remnants, which can both be seen in the spectrum at phase 14.010 in Fig.~\ref{fig 1: eta Car All}. {We attempted to measure the two features at $-145$ and $-168$ \kms\ independently, but the features were blended and weak and non-Gaussian in nature. The resulting measurements were deemed unreliable by both their component errors and visual inspection of the spectra compared with the measurements.}

As seen in Fig. \ref{fig 4: $-$168 \kms}, there is variability in $W_{\lambda}$ of the $-$168 \kms\ feature. This originates in a decreasing level of ionization near periastron. The increased activity near $\phi$ = 14.00 resulted in the steady climb in magnitude of $W_{\lambda}$, which then begins to decrease near $\phi$ = 14.018. This process is similar to that of the $-$145 \kms\ feature. The timing of the disappearance of the $-$168 \kms\ feature was interrupted by the temporary closure of the facilities at CTIO.

\subsection{The Absorption Component at +87 \kms}
\begin{figure}[t!]
    \centering
    \includegraphics[width=\columnwidth]{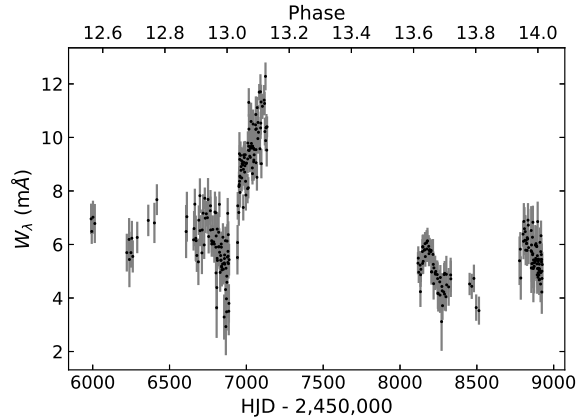}
    \caption{The measured equivalent width strength of the $+$87 \kms\ feature from the CHIRON spectrograph, where the strength changes slightly near the 2014 periastron, presumably because of the changing profile of the occulter.}
    \label{fig 5: +87}
\end{figure}

Following telluric correction, a previously-unnoticed absorption component was recognized at $\sim$+87 \kms. This component, seen in Figure \ref{fig 1: eta Car All}, indicates that material is moving towards $\eta$ Car and would represent an inflow of material if originating in the gas surrounding $\eta$ Car. A second, identical component at $-217$ \kms\ in the {Na D}$_1$ reference frame is also observed and represents the {Na D}$_2$ transition at the same velocity. 

Equivalent width measurements for the +87 \kms\ feature were taken between the wavelengths of
5897.55{\AA} and 5897.74{\AA}, corresponding to the radial velocities of 78.87 \kms\ and 88.53 \kms and are shown in Fig.~\ref{fig 5: +87}. Each measurement was made within the bounds of .19{\AA}, or 9.66 \kms. We notice a small change around the 2014 periastron passage, which may be explained by the change in the $V$-band brightness across periastron such as that observed by \citet{2019MNRAS.484.1325D}. In this regard, we see an increase in continuum flux leading up to the event, followed by a dip in the flux after periastron. With the equivalent width depending on the continuum, as shown by the associated equivalent width equation earlier in this section, we see that the small differences at that time can be explained by the change in flux. Furthermore, oscillations at times away from periastron may reflect changes in the photometric brightness, such as those recently documented by \citet{2018MNRAS.475.5417R}. However, as noted by \citet{2010AJ....139.1534R}, usually changes in the equivalent widths due to the continuum are primarily a concern for large flux lines such as the H$\alpha$ line. It is also possible that these small changes are due to changes in the {Na D} emission in the region of this absorption component, which could alter our measurements. The exact amount of variation that could be intrinsic to this component should therefore be very small when compared to the large changes in the underlying flux from the system along with the variability of the broad Na D wind emission features present in this region of the spectrum.

\section{Discussion}

\begin{figure*}[t!]
    \centering
    \includegraphics[width=\textwidth]{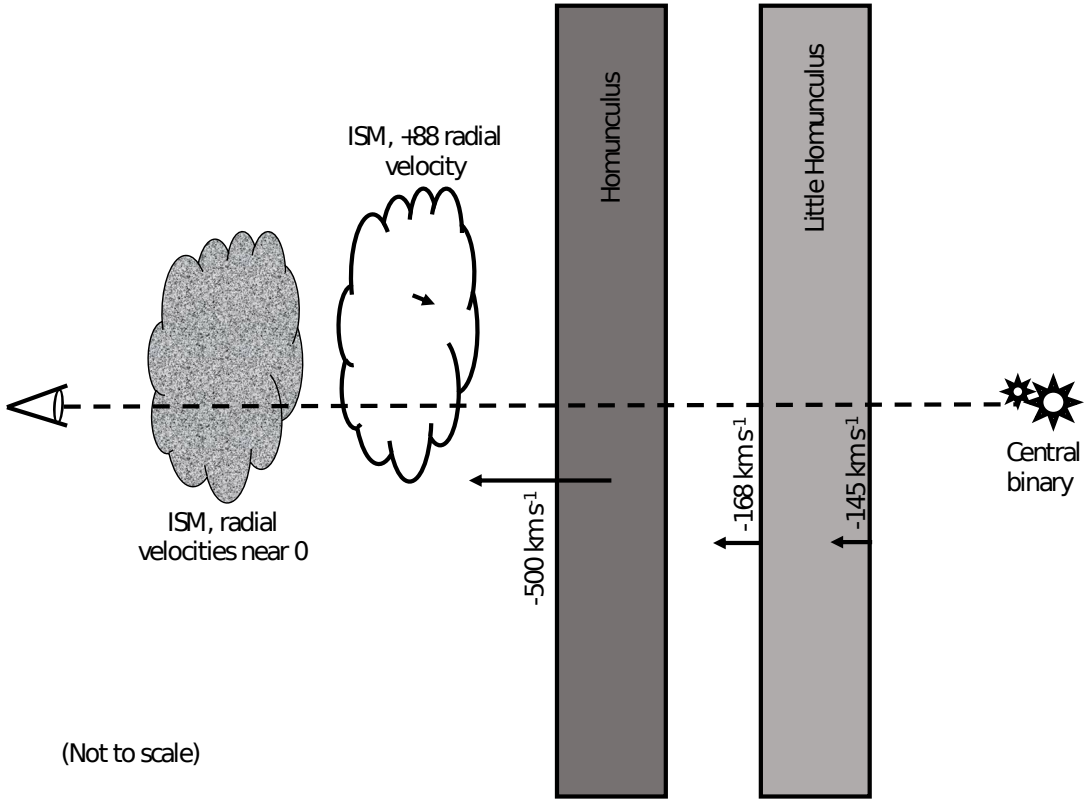}
    \caption{A cartoon diagram showing the geometry of the gas that provides {Na D} absorption. A detailed discussion of this geometry is in the text. }
    \label{fig 6: cartoon}
\end{figure*}

Our measurements of {Na D}$_1$ absorption components from $\eta$ Carinae can be used to illuminate some recent changes in the spectrum and photometric record. In particular, \citet{2019MNRAS.484.1325D, 2021MNRAS.505..963D} found that there are long-term changes related to the circumstellar extinction in the line of sight that should conclude {in $\sim 2032$. Our results show that the Na D absorption from the Little Homunculus is much weaker and likely absent at times away from periastron, in broad agreement with the results from \citet{2019MNRAS.484.1325D}.} The spectral modeling of the system, first accomplished with {\it HST} spectroscopy and the CMFGEN non-LTE radiative transfer code by \citet{1998ApJ...496..407H, 2001ApJ...553..837H}, required 2.0 magnitudes of grey extinction in the line of sight to model both the stellar (wind) spectrum and achieve the correct line strengths for the narrow line emission from the surrounding Weigelt knots, but also had 5 magnitudes of extinction in $V$-band in the line of sight towards the binary. We show a cartoon diagram of the most likely geometry that could accommodate the observations in Fig.~\ref{fig 6: cartoon}.

If we assume that the component near $-$145 \kms\ originates in the Little Homunculus as noted in several studies \citep[e.g.,][]{1982A&A...111..375M, 1998A&AS..133..299D, 2005AJ....129..900D, 2005A&A...435..183J, 2006ApJS..163..173G, 2010AJ....139.1534R, 2015AJ....150..109R}, then we can begin to interpret the variability and character of the components at $-$145 and $-$168 \kms. First, we consider the variability of the component at $-$145 \kms. \citet{2010AJ....139.1534R} describe the variability of the component at the same velocity for the H$\alpha$ transition. For H$\alpha$, \citet{2010AJ....139.1534R} postulate that during the spectroscopic minimum near periastron, the secondary star would pass behind the primary, and thus drop the ionizing radiation in our line of sight. This drop in ionization would then relax the gas and allow lower ionization transitions to be observed in absorption.

These same principles can explain the changes we see in both the $-$145 \kms\ component (Fig.~\ref{fig 2: $-$145 \kms}) across the 2009 and 2014 periastron passages and the $-$168 \kms\ component (Fig.~\ref{fig 4: $-$168 \kms}) during the recent 2020 periastron event. The $-$168 \kms\ absorption was only previously observed in six low-ionization lines (Mg I, Mg II, Cr II, Mn II, Fe II, and Ni II) in the near ultraviolet by \citet{2006ApJS..163..173G} and only during the low-ionization state of the binary. In contrast, the $-$145 \kms\ was observed in the same lines, as well as in He I, Ti II, V II, Co II, and Ni II, {although \citet{2006ApJS..163..173G} reported it as $-146$ \kms}.

\begin{figure*}
\includegraphics[width=18 cm]{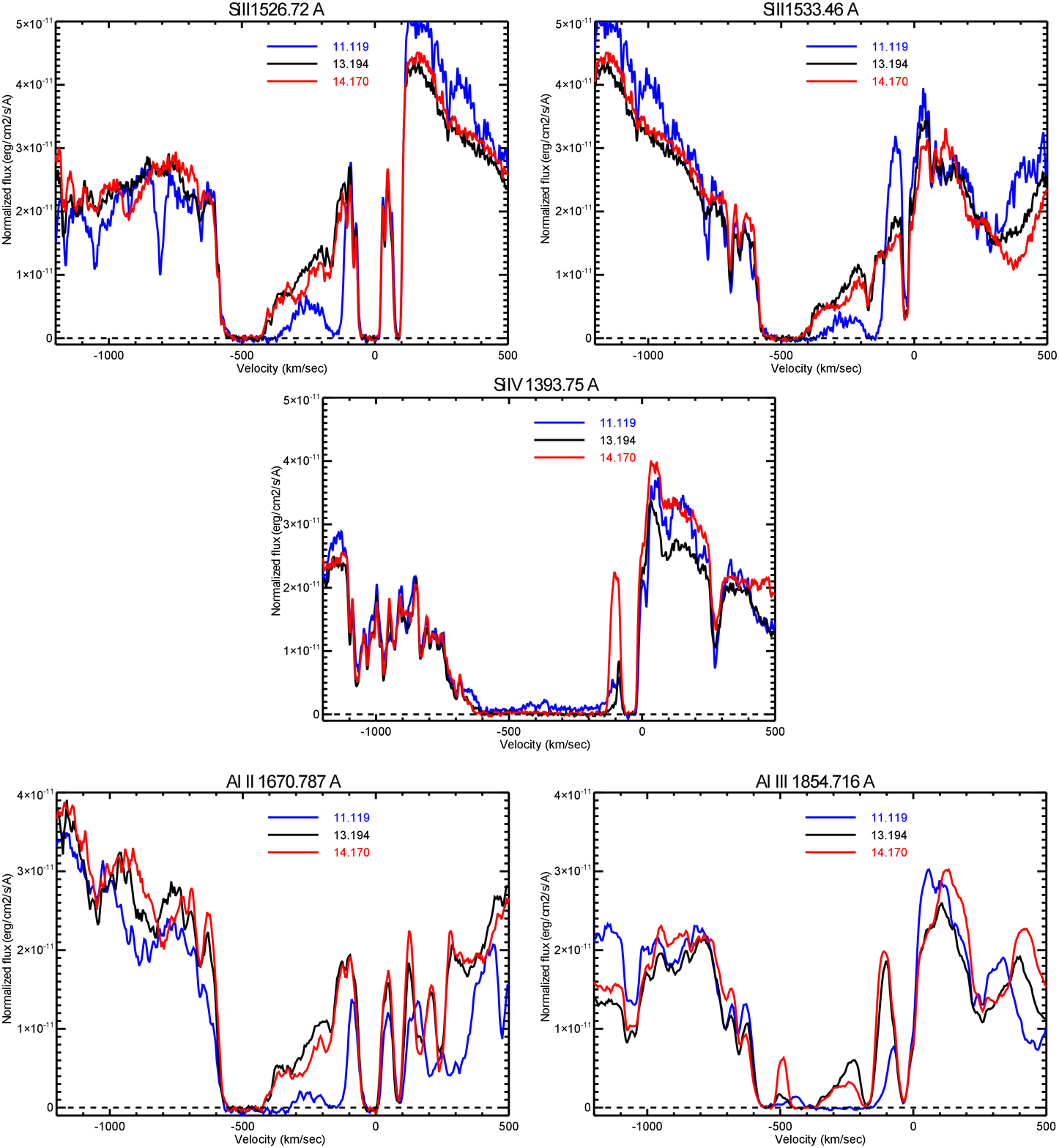}
\caption{Velocity profiles of \ion{Si}{2}, \ion{Si}{4}, \ion{Al}{2} and \ion{Al}{3} resonant lines as recorded by {\it HST}/STIS. The three spectra were recorded at similar phases following periastron passages 11, 13 and 14 when $\eta$ Carinae has re-entered the high-ionizaton state. Long-term changes in the Little Homunculus are apparent between $-$100 and $-$400 \kms. The absorption at $+$87 \kms\ is also present in the low ionization Si II and Al II profiles.}
\label{fig 7: FUV}
\end{figure*}

In the far-ultraviolet, several transitions have been observed by {\it HST}/STIS, allowing us to make comparisons to the {Na D}$_1$ observations described here. For this discussion, we limit ourselves to the \ion{Al}{2} $\lambda$1671, \ion{Al}{3} $\lambda$1855, \ion{Si}{2} $\lambda$1527, and \ion{Si}{4} $\lambda$1394 transitions {which were observed during the previous two periastron passages where we also have CHIRON spectroscopy}. All of these {ultraviolet} transitions and {Na D} are resonance lines. In Fig.~\ref{fig 7: FUV}, we show these transitions at phases 11.119, 13.194, and 14.170 so that the binary-induced phase differences are minimal, and we can examine the long-term evolution of the profiles. Each profile was normalized to the `continuum' flux level near 1483\AA\ recorded at phase $\phi =$ 14.170. Gull et al. (2022, in prep) demonstrate that normalization of the `continuum' is valid since virtually all continuum originates from deep within $\eta$ Car-A.

These profiles show that these species have evolved to show less absorption at $-$145 \kms, {although these changes are amplified because the normalization of these data}. This shows a similar trend as the {Na D} observations, where recent observations since phase $\sim$13.5 have had no absorption present at that velocity except for a brief appearance at the cycle 14 periastron event. During the recent periastron, this absorption was seen at higher velocity and with a companion absorption component at $-$168 \kms. In the ultraviolet profiles, the lower ionization \ion{Al}{2} and \ion{Si}{2} lines show that the absorption components at these velocities are also seen at higher velocities than in earlier cycles, such as just following the cycle 11 periastron passage. 

We postulate that the absorption component was formed in a spatially unresolved dense clump within the Little Homunculus. This clump would respond to the ionization changes induced by the companion's orbit around the primary star. In this situation, the clump would be either moving out of our line of sight or dissipating with time. Both scenarios allow for a weaker absorption profile now than in the past, consistent with the observations of the ultraviolet resonance lines and the {Na D} profiles.

We used a curve of growth method to find the column density of the absorbing Na atoms during the times of just prior to the 2009 event {(phases near 11.95)}, near phase 12.5, and after the 2014 event at phase 13.1. We used these times to avoid times when the ionization changes, which would therefore change the column density of neutral sodium in our line of sight. At these three epochs, the equivalent widths were roughly 70$\pm12$ m\AA, 52$\pm$2.5 m\AA, and 14.7$\pm$2.2 m\AA\ respectively. With the known oscillator strength and wavelength of Na D$_1$, we can then calculate the values of $W_\lambda / \lambda$ to obtain the values of $F(\tau)$ using the curve of growth from \citet{Spitzer} with an assumed line width of 10 \kms, which is complimentary to techniques used for the local interstellar medium \citep{1990ApJ...358..473W}. Then, we derive a column density of $4.9\times 10^{11}$ cm$^{-2}$, $3.5\times 10^{11}$ cm$^{-2}$, and $0.8\times 10^{11}$ cm$^{-2}$ for these three epochs respectively, confirming the lower column density in recent times. This is in agreement with the changes in the ultraviolet flux which has increased by about a factor of 10 in the same time period \citep{2021ApJ...923..102G}.

\begin{figure*}[t!]
    \centering
    \includegraphics[width=6.5in]{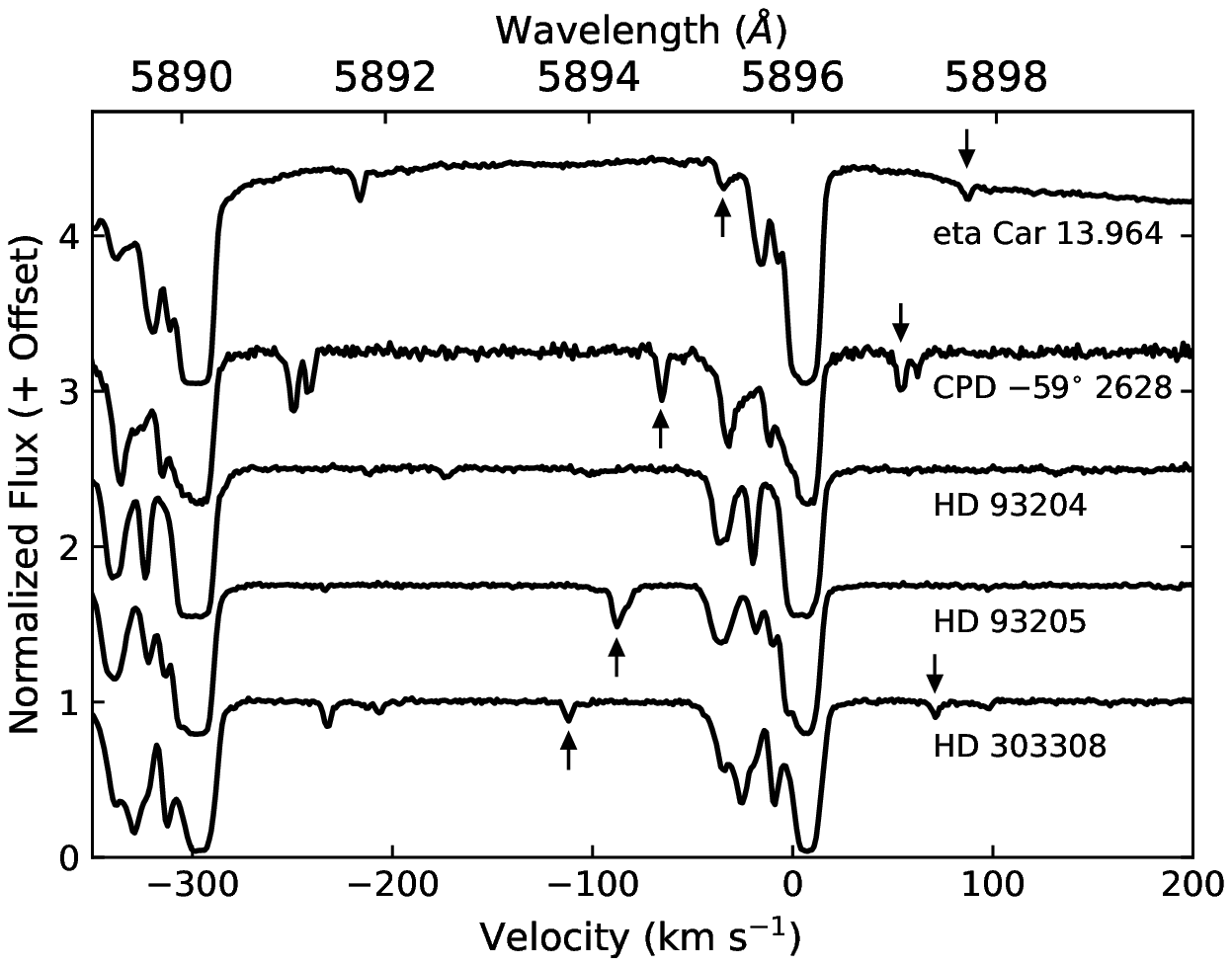}
    \caption{A comparison of the {Na D} profiles of $\eta$ Carinae at $\phi = 13.964$, CPD -59$^{\circ}$ 2628, HD 93204, HD 93205, and HD 303308 at 1 arcmin to north. $\eta$ Carinae, CPD -59$^{\circ}$ 2628, and HD 303308 clearly have features present at positive velocities similar to that of $\eta$ Car. {Similar interstellar features are denoted with arrows.}}
    \label{fig 8: HD Star Comparison}
\end{figure*}

With a clump in the line of sight, we can also imagine that the angular size was relatively small to us. The neighboring Weigelt knots, separated from the central source by only about 0.2\arcsec \citep{1986A&A...163L...5W}, did not undergo the same amount of extinction as the central source according to the models of \citet{2001ApJ...553..837H}. As a result, an angularly small clump {similar to the Weigelt knots, either} within the Little Homunculus {or in our line of sight} could provide the necessary extinction to provide the absorption lines and the added extinction toward the binary. As \citet{2019MNRAS.484.1325D, 2021MNRAS.505..963D} have noted, the central occulter must be dissipating to explain the object's brightening and the lack of observed changes in the object's spectrum. As a result, the simplicity of a dissipating or moving clump in our line of sight would help explain many of the unusual observations of this system, {although neither explanation can be fully refuted with the current data}.

The component at $-$145 \kms\ is also seen to shift in its velocity during the last several cycles as it was dissipating. The movement of the clump out of the line-of-sight or its thicker parts dissipating on the interior of the Little Homunculus shell, as illustrated in our Fig.~\ref{fig 6: cartoon}, would allow for an apparent shift in the radial velocity of the component by having the densest part of the clump being at different points in the outflow of the Little Homunculus. Furthermore, as the thickest part of the clump at the central velocity of the clump dissipated or moved out of the line of sight, the changing ionization front would penetrate into the Little Homunculus allowing for the {weaker} $-$168 \kms\ to be visible during the most recent periastron passage while not observed in {Na D} in past events.

Moreover, \citet{2016MNRAS.462.3196G} recently reported on spatially resolved {\it HST}/STIS spectroscopy of the binary across an entire orbit of the system. The results indicate that the Weigelt knot B \citep[A4 in the analysis of ][]{1986A&A...163L...5W} has fully dissipated or disappeared in the surrounding region of $\eta$ Carinae. {If the clump is similar ejecta as the Weigelt knots that were ejected during the Lesser Eruption, then the clump} is traversing the region rather than dissipating, then it could now be providing additional extinction toward this clump. However, Weigelt B could also be dissipating with time, meaning that the clumps left over from the Lesser Eruption of $\eta$ Car could all be beginning to dissipate around this time, as the Weigelt knots have been shown to be associated with the Lesser Eruption of $\eta$ Carinae \citep{2012ASSL..384..129W}.

The origin of the component at $+$87 \kms\ is simpler to explain. With an apparent lack of variability (Fig.~\ref{fig 5: +87}), it is hard to envision a situation in which that component would actually represent inflowing material from the surrounding winds and ejecta. We anticipated this being interstellar in origin so we tested our hypothesis in two ways. First, we see in the low-ionization lines of \ion{Al}{2} and \ion{Si}{2} that there is a component at roughly the same velocity as the $+$87 \kms\ component of {Na D}$_1$ and {Na D}$_2$. This component is not seen to vary in its absolute absorption with the {\it HST} observations recorded over an interval of 16.5 years (Fig.~\ref{fig 7: FUV}). The $+$87 \kms\ component is not present in \ion{Si}{4} $\lambda$1394, \ion{Al}{3} $\lambda$1855 nor in \ion{Si}{2} $\lambda$1533. The absorbing cloud contains only singly-ionized ions and is sufficiently cool that no \ion{Si}{2} ions populate the low-lying energy level at 287 cm$^{-1}$.

These features were reminiscent of the ultraviolet lines of other stars in the Carina Nebula complex observed by \citet{2002ApJS..140..407W}. Strong absorption has been previously observed in the nearby CPD -59$^{\circ}$2603, as noted in \citet{2001ApJ...547L.155D}. Small line widths of high-velocity features, as well as column density variations, indicate structures with transverse velocities when compared to radial velocities. \citet{2002ApJS..140..407W} depicts these absorptions from both 1997 and 1999 observations. These data show, in total, nine strong, low-ionization absorptions within our line of sight. The strongest features appear in lower-velocity regions, but it should be noted that dominant components of high-ionization states appear at intermediate and negative velocities. The high-ionization absorption corresponds to the global expansion of the H II region. This expansion is further investigated in \citet{2007PASP..119..156W}. 
These observations of neutral sodium at varying velocities for stars in the Carina Nebula raise the possibility that they are formed in low-density outflowing gas from the region, perhaps originating from a past supernova {or the collective winds of the cluster}.

 To further test the interstellar origin of this component, we observed three of the Carina stars observed by \citet{2002ApJS..140..407W} with the CHIRON spectrograph at the same resolving power as our $\eta$ Carinae observations, along with CPD -59$^{\circ}$ 2628 which is very close to $\eta$ Car in the sky\footnote{CPD -59$^{\circ}$ 2628 was seen to interfere with $\eta$ Car's photometric light curve observed by {\it BRITE-Constellation} reported by \citet{2018MNRAS.475.5417R} due to its close proximity in the sky.}. For these observations of {Na D} shown in Fig.~\ref{fig 8: HD Star Comparison}, we see many similarities in the interstellar lines. 

The primary interstellar lines near 0 \kms\ all show similar components for these stars. As noted by \citet{2002ApJS..140..407W}, many of these stars have lines at high velocities. The {Na D} components of the neighboring CPD -59$^{\circ}$2603 show a positive velocity component at lower velocity than that of $\eta$ Carinae, with a similar profile showing a double absorption. The observations of $\eta$ Car show a similar double component of absorption. These similarities with other stars within the Carina Nebula show that they could be explained by a {complicated} foreground absorption complex, perhaps from a past supernova or cluster wind that shows complexity across the region, {but the physical cause for these complex interstellar absorptions is beyond the scope of this paper}.

\section{Conclusions}

Our study of the {Na D} complex of $\eta$ Carinae has found several interesting trends that relate to other observations of the system. 

\begin{itemize}
\item{The absorption component near $-$145 \kms\ has been observed to move to higher negative velocities over the past decade. Simultaneously, the strength of this absorption has weakened overall, becoming indiscernible by 2018. This happened in parallel to the brightening of the star while the nebula has remained more consistent in its brightness. This points to the central occulter being a spatially unresolved clump in the Little Homunculus.}
\item{We find a new absorption in the spectra that is near $+$87 \kms\, which seems to be interstellar in origin. Similarly shaped structures are seen in neighboring stars, but at different velocities. The variability of this component can likely be attributed to changes in the continuum flux and Na D wind emission during the periastron passages. }
\item{The interstellar feature at $+$87 \kms\ is also seen in some low-ionization ultraviolet resonance lines. }
\item{To further disentangle if the occulter is/was in the Little Homunculus or another clump interior to the Little Homunculus, future observations of the Na profile at high resolution and in reflected light from various lines of sight in the Homunculus could illuminate this discussion.}

\end{itemize}

\section*{Acknowledgements}

Many measurements in this project were originally made by Lucas St-Jean, and we thank him for these early contributions that led to these results. The CTIO observations are the result of many allocations of telescope time. We thank internal SMARTS allocations at Georgia State University (in 2008--2009), as well as NOIR Lab (formerly NOAO) allocations of NOAO-09B-153, NOAO-12A-216, NOAO-12B-194, NOAO-13B-328, NOAO-15A-0109, NOAO-18A-0295, NOAO-19B-204, NOIRLab-20A-0054, and NOIRLab-21B-0334. C.S.P. and A.L. were partially supported by the Embry-Riddle Aeronautical University Undergraduate Research Institute. E.S. acknowledges support from the Arizona Space Grant program. N.D.R., C.S.P., A.L., E.S., and T.R.G. acknowledge support from the {\it HST} GO Programs \#15611 and \#15992. A.F.J.M. is grateful to NSERC (Canada) for financial aid.


\bibliography{sample631}{}
\bibliographystyle{aasjournal}

\clearpage

\appendix
\section{Online Tables}
\startlongtable
\begin{deluxetable}{ccc}
\tablecaption{$-$ 145 \kms Measurements \label{table:145}}
\tablewidth{0pt}
\centering
\tablehead{
  \colhead{HJD}         &
  \colhead{$W_{\lambda}$}  &
  \colhead{$\sigma$}  \\  
  \colhead{-2,450,000}         &
  \colhead{}  &  
  \colhead{}         \\ 
  \colhead{(d)}  &
  \colhead{(m\AA)}  &
  \colhead{(m\AA)}
  }
\startdata
4755.8605 & 74.4100  & 6.0000 \\
4784.8577 & 56.6100  & 6.0000 \\
4818.8060 & 64.8500  & 10.7695 \\
4819.8108 & 72.8200  & 7.6075 \\
4821.8520 & 73.1500  & 6.1266 \\
4824.7536 & 70.4100  & 4.5138 \\
4829.7838 & 72.7000  & 6.0701 \\
4831.7530 & 90.4600  & 4.2639 \\
4836.7382 & 127.0000 & 5.5035 \\
4836.7510 & 92.0700  & 5.6323 \\
4837.7318 & 106.9000 & 3.6945 \\
4838.7569 & 125.0000 & 3.4637 \\
4839.7867 & 100.1000 & 3.3596 \\
4840.7789 & 107.0000 & 3.2727 \\
4841.7684 & 130.0000 & 4.6655 \\
4842.7738 & 126.2000 & 2.6296 \\
4843.7735 & 115.9000 & 2.8100 \\
4844.7762 & 155.2000 & 3.0682 \\
4845.7651 & 158.0000 & 3.4639 \\
4846.7704 & 160.0000 & 3.7042 \\
4847.7971 & 121.9000 & 4.0158 \\
4848.7409 & 176.1000 & 5.2701 \\
4849.7256 & 186.4000 & 4.1598 \\
4850.7334 & 142.9000 & 3.0279 \\
4851.7190 & 216.0000 & 4.8571 \\
4852.7420 & 205.5000 & 4.6979 \\
4853.7283 & 221.2000 & 5.3942 \\
4854.7404 & 187.9000 & 4.5506 \\
4855.7358 & 185.4000 & 3.6463 \\
4857.7856 & 194.1000 & 3.2556 \\
4858.7676 & 186.0000 & 3.2942 \\
4859.7749 & 177.8000 & 3.8427 \\
4861.7267 & 182.4000 & 2.7894 \\
4862.7393 & 176.2000 & 4.8369 \\
4863.7592 & 189.5000 & 2.4151 \\
4864.7284 & 193.9000 & 10.1812 \\
4865.7026 & 180.9000 & 4.0439 \\
4866.7166 & 183.4000 & 3.5817 \\
4867.7075 & 186.9000 & 3.2210 \\
4875.6516 & 179.8000 & 4.0514 \\
4879.6606 & 193.0000 & 4.7157 \\
4880.6719 & 165.2000 & 5.1300 \\
4881.6680 & 161.3000 & 6.8569 \\
5125.8682 & 68.2600  & 6.0000 \\
5139.8436 & 70.3600  & 6.0000 \\
5140.8246 & 82.1900  & 6.0000 \\
5164.8118 & 68.4000  & 6.0000 \\
5213.7590 & 79.0900  & 6.7142 \\
5226.7402 & 57.2000  & 4.0890 \\
5230.6255 & 70.7500  & 6.0000 \\
5240.6474 & 64.5900  & 4.7343 \\
5250.6127 & 64.4800  & 3.0098 \\
5254.5753 & 63.3300  & 3.6512 \\
5259.6044 & 65.3400  & 4.1624 \\
5270.5564 & 64.5900  & 3.2782 \\
5278.4984 & 74.3400  & 4.6157 \\
5284.5346 & 66.7000  & 3.5030 \\
5293.5099 & 69.6000  & 3.0000 \\
5306.5008 & 71.9600  & 2.4000 \\
5311.5186 & 84.3400  & 1.8000 \\
5314.5970 & 65.2800  & 1.2000 \\
5317.4860 & 64.8600  & 1.5000 \\
5327.4623 & 65.0300  & 1.0000 \\
5335.4752 & 65.0700  & 1.0000 \\
5360.3720 & 67.4400  & 1.2147 \\
5989.6915 & 53.8782  & 1.2147 \\
5993.7546 & 49.2825  & 1.2244 \\
6001.7288 & 53.9522  & 1.2145 \\
6013.6892 & 50.7330  & 1.2213 \\
6221.8876 & 45.9027  & 1.2315 \\
6236.8436 & 45.9535  & 1.2314 \\
6238.8674 & 47.0313  & 1.2291 \\
6248.8123 & 41.3750  & 1.2410 \\
6254.8795 & 41.8054  & 1.2401 \\
6260.7808 & 45.2798  & 1.2328 \\
6289.7971 & 42.6745  & 1.2383 \\
6361.6472 & 42.3128  & 1.2391 \\
6401.5905 & 42.9964  & 1.2376 \\
6417.5924 & 39.5673  & 1.2449 \\
6607.8431 & 37.8184  & 1.2485 \\
6612.8653 & 45.5610  & 1.2322 \\
6656.7032 & 47.6766  & 1.2278 \\
6659.7093 & 45.8957  & 1.2315 \\
6664.6745 & 49.2717  & 1.2244 \\
6670.7826 & 48.1998  & 1.2267 \\
6672.8378 & 50.6189  & 1.2216 \\
7038.7462 & 27.4097  & 1.2705 \\
6677.7595 & 43.8922  & 1.2357 \\
6687.7085 & 45.7443  & 1.2318 \\
6690.6917 & 44.6191  & 1.2342 \\
6697.7271 & 45.8772  & 1.2316 \\
6698.7748 & 46.1766  & 1.2309 \\
6710.6697 & 44.4810  & 1.2345 \\
6712.6945 & 47.4635  & 1.2282 \\
6718.7120 & 47.7831  & 1.2275 \\
6725.5713 & 49.0575  & 1.2249 \\
6729.5623 & 44.4447  & 1.2346 \\
6732.6062 & 44.0142  & 1.2355 \\
6739.5490 & 44.0139  & 1.2355 \\
6746.5203 & 42.7372  & 1.2382 \\
6750.5418 & 45.3872  & 1.2326 \\
6754.5540 & 44.7233  & 1.2340 \\
6765.5903 & 44.2191  & 1.2351 \\
6766.5463 & 44.7315  & 1.2340 \\
6767.5116 & 44.4612  & 1.2345 \\
6774.5470 & 40.5219  & 1.2428 \\
6775.4705 & 42.4427  & 1.2388 \\
6777.5390 & 43.6835  & 1.2362 \\
6781.4742 & 40.9965  & 1.2418 \\
6786.5036 & 43.9414  & 1.2356 \\
6787.5061 & 42.9262  & 1.2378 \\
6791.5527 & 48.3870  & 1.2263 \\
6795.5095 & 47.9230  & 1.2272 \\
6800.4644 & 44.7013  & 1.2340 \\
6801.5000 & 44.1911  & 1.2351 \\
6802.5067 & 45.2528  & 1.2329 \\
6803.5294 & 48.5563  & 1.2259 \\
6806.5318 & 44.8490  & 1.2337 \\
6807.6147 & 45.1985  & 1.2330 \\
6809.4604 & 48.2952  & 1.2265 \\
6810.4839 & 48.4599  & 1.2261 \\
6811.4890 & 50.8294  & 1.2211 \\
6818.5059 & 48.9579  & 1.2251 \\
6819.5294 & 52.8926  & 1.2168 \\
6823.5395 & 48.1640  & 1.2267 \\
6824.5183 & 50.8258  & 1.2211 \\
6825.5069 & 49.0499  & 1.2249 \\
6829.4942 & 52.4642  & 1.2177 \\
6832.5015 & 52.2552  & 1.2181 \\
6833.5015 & 51.0414  & 1.2207 \\
6835.5347 & 51.0836  & 1.2206 \\
6836.4977 & 51.9177  & 1.2188 \\
6837.4984 & 49.9543  & 1.2230 \\
6838.5142 & 51.2240  & 1.2203 \\
6839.5007 & 50.5233  & 1.2218 \\
6840.4992 & 50.1879  & 1.2225 \\
6845.4618 & 53.6993  & 1.2151 \\
6847.4567 & 54.8509  & 1.2126 \\
6850.5099 & 54.4124  & 1.2136 \\
6855.4628 & 57.0751  & 1.2079 \\
6856.4542 & 59.2667  & 1.2033 \\
6857.4572 & 62.5979  & 1.1963 \\
6858.4560 & 64.9401  & 1.1914 \\
6859.4699 & 67.0792  & 1.1869 \\
6863.4696 & 85.0692  & 1.1489 \\
6864.4670 & 93.0745  & 1.1320 \\
6866.5170 & 105.4488 & 1.1060 \\
6867.4637 & 118.6766 & 1.0781 \\
6870.4592 & 133.8580 & 1.0461 \\
6871.4663 & 137.0995 & 1.0392 \\
6872.4634 & 150.3551 & 1.0113 \\
6873.4627 & 148.4821 & 1.0152 \\
6874.4542 & 155.9589 & 0.9995 \\
6876.4581 & 151.2320 & 1.0094 \\
6878.4816 & 155.9124 & 0.9996 \\
6879.4665 & 163.0676 & 0.9845 \\
6880.4665 & 166.2434 & 0.9778 \\
6881.4646 & 162.9879 & 0.9846 \\
6882.4610 & 163.5481 & 0.9835 \\
6883.4684 & 163.5807 & 0.9834 \\
6885.4659 & 156.7293 & 0.9978 \\
6886.4726 & 155.3184 & 1.0008 \\
6887.4599 & 148.2345 & 1.0157 \\
6941.9068 & 83.2774  & 1.1527 \\
6942.9073 & 78.9372  & 1.1619 \\
6943.9038 & 79.2119  & 1.1613 \\
6950.8754 & 65.4337  & 1.1903 \\
6951.8733 & 61.1214  & 1.1994 \\
6952.8645 & 57.7236  & 1.2066 \\
6953.8796 & 56.6173  & 1.2089 \\
6954.8312 & 55.2559  & 1.2118 \\
6955.8613 & 54.3239  & 1.2137 \\
6956.8738 & 51.8092  & 1.2190 \\
6957.8902 & 49.4740  & 1.2240 \\
6958.8639 & 47.6115  & 1.2279 \\
6959.8471 & 47.2173  & 1.2287 \\
6961.8481 & 47.7860  & 1.2275 \\
6964.8650 & 46.2104  & 1.2309 \\
6965.8623 & 43.0701  & 1.2375 \\
6968.8670 & 42.6636  & 1.2383 \\
6969.8629 & 46.9872  & 1.2292 \\
6972.8807 & 44.0054  & 1.2355 \\
6974.8472 & 39.0575  & 1.2459 \\
6977.8456 & 42.3540  & 1.2390 \\
6979.8393 & 40.8341  & 1.2422 \\
6980.8055 & 44.9754  & 1.2335 \\
6982.8125 & 40.4891  & 1.2429 \\
6984.8198 & 37.2191  & 1.2498 \\
6985.8277 & 39.8160  & 1.2443 \\
6986.8716 & 36.3258  & 1.2517 \\
6988.7943 & 35.3878  & 1.2537 \\
6990.7638 & 36.4655  & 1.2514 \\
6992.7814 & 38.7161  & 1.2467 \\
6994.8169 & 37.5105  & 1.2492 \\
6996.8516 & 35.9668  & 1.2525 \\
6998.8053 & 37.3780  & 1.2495 \\
7000.8371 & 39.6315  & 1.2447 \\
7007.8533 & 36.3848  & 1.2516 \\
7008.7387 & 33.2607  & 1.2582 \\
7009.7487 & 36.4337  & 1.2515 \\
7012.8169 & 32.8265  & 1.2591 \\
7013.8113 & 30.2994  & 1.2644 \\
7014.7384 & 28.1325  & 1.2690 \\
7015.7661 & 29.0392  & 1.2671 \\
7016.7963 & 33.6223  & 1.2574 \\
7018.7787 & 30.2510  & 1.2645 \\
7021.8358 & 28.3041  & 1.2686 \\
7022.7519 & 28.4105  & 1.2684 \\
7025.8280 & 26.3324  & 1.2728 \\
7026.7004 & 26.4958  & 1.2724 \\
7030.6911 & 29.6624  & 1.2657 \\
7032.8183 & 27.5399  & 1.2702 \\
7035.7243 & 27.2459  & 1.2708 \\
7037.8164 & 27.6261  & 1.2700 \\
7038.8521 & 29.9518  & 1.2651 \\
7039.8336 & 26.5309  & 1.2723 \\
7040.7121 & 26.6035  & 1.2722 \\
7043.8463 & 27.2521  & 1.2708 \\
7046.6847 & 28.0830  & 1.2691 \\
7047.6855 & 27.1237  & 1.2711 \\
7048.8103 & 25.4844  & 1.2746 \\
7050.7318 & 23.4698  & 1.2788 \\
7051.8465 & 25.7026  & 1.2741 \\
7059.7902 & 20.4370  & 1.2852 \\
7060.6626 & 26.0203  & 1.2734 \\
7061.8510 & 24.8567  & 1.2759 \\
7062.6967 & 22.0224  & 1.2819 \\
7063.7354 & 21.3341  & 1.2833 \\
7064.7755 & 22.7382  & 1.2803 \\
7068.7710 & 23.0571  & 1.2797 \\
7070.6468 & 21.0678  & 1.2839 \\
7074.8003 & 19.7124  & 1.2867 \\
7075.6999 & 17.5518  & 1.2913 \\
7078.8993 & 17.4877  & 1.2914 \\
7083.6618 & 18.0506  & 1.2902 \\
7088.6417 & 13.9511  & 1.2989 \\
7091.5928 & 18.4445  & 1.2894 \\
7092.7233 & 12.8638  & 1.3012 \\
7093.7108 & 13.2399  & 1.3004 \\
7095.8005 & 14.4630  & 1.2978 \\
7097.6962 & 11.1866  & 1.3047 \\
7104.5614 & 13.8121  & 1.2992 \\
7115.7489 & 15.5300  & 1.2955 \\
7118.5573 & 15.6659  & 1.2953 \\
7120.5391 & 15.9264  & 1.2947 \\
7124.4992 & 11.0033  & 1.3051 \\
7125.5235 & 13.5067  & 1.2998 \\
7127.6192 & 16.6063  & 1.2933 \\
7130.5443 & 13.2235  & 1.3004 \\
7132.5158 & 15.0305  & 1.2966 \\
7134.6264 & 14.2428  & 1.2983
\enddata
\end{deluxetable}

\startlongtable
\begin{deluxetable}{ccc}
\tablecaption{$+$ 87 \kms Measurements \label{table:87}}
\tablewidth{0pt}
\centering
\tablehead{
  \colhead{HJD}         &
  \colhead{$W_{\lambda}$}  &
  \colhead{$\sigma$}  \\  
  \colhead{-2,450,000}         &
  \colhead{}  &  
  \colhead{}         \\ 
  \colhead{(d)}  &
  \colhead{(m\AA)}  &
  \colhead{(m\AA)}
  }
\startdata
5989.6915 & 6.9575  & 0.5690 \\
5993.7546 & 6.4877  & 0.4666 \\
6001.7288 & 7.0189  & 0.6171 \\
6013.6892 & 6.7837  & 0.7300 \\
6221.8876 & 5.6955  & 0.7066 \\
6236.8436 & 6.1888  & 0.7984 \\
6238.8674 & 5.4392  & 1.0361 \\
6248.8123 & 5.6955  & 0.5610 \\
6254.8795 & 6.2337  & 0.6488 \\
6260.7808 & 5.5529  & 0.6005 \\
6289.7971 & 6.2622  & 0.5882 \\
6361.6472 & 6.8988  & 0.5884 \\
6401.5905 & 6.8054  & 0.7523 \\
6417.5924 & 7.6743  & 0.5769 \\
6607.8431 & 6.4830  & 0.9925 \\
6612.8653 & 7.0374  & 0.9321 \\
6656.7032 & 6.1734  & 0.4011 \\
6659.7093 & 6.5938  & 0.6570 \\
6664.6745 & 7.5062  & 0.5782 \\
6670.7826 & 6.2313  & 0.8487 \\
6672.8378 & 6.1406  & 0.7884 \\
7038.7462 & 8.8609  & 0.4875 \\
6677.7595 & 5.5892  & 0.6148 \\
6687.7085 & 5.3510  & 0.8682 \\
6690.6917 & 6.9078  & 0.5528 \\
6697.7271 & 7.8226  & 0.6497 \\
6698.7748 & 6.5161  & 0.5798 \\
6710.6697 & 5.6846  & 0.7924 \\
6712.6945 & 6.5423  & 0.7655 \\
6718.7120 & 7.1689  & 0.7073 \\
6725.5713 & 5.9835  & 0.5270 \\
6729.5623 & 7.7273  & 0.4005 \\
6732.6062 & 6.9829  & 0.5528 \\
6739.5490 & 7.1471  & 0.4769 \\
6746.5203 & 7.0010  & 0.5962 \\
6750.5418 & 7.6828  & 0.7995 \\
6754.5540 & 6.2975  & 0.6330 \\
6765.5903 & 7.2603  & 0.5293 \\
6766.5463 & 7.4462  & 0.5885 \\
6767.5116 & 6.5413  & 0.5541 \\
6774.5470 & 6.1398  & 0.5976 \\
6775.4705 & 6.0161  & 0.3171 \\
6777.5390 & 6.1015  & 0.3593 \\
6781.4742 & 6.5604  & 0.5276 \\
6786.5036 & 6.0340  & 0.4806 \\
6787.5061 & 6.0957  & 0.5503 \\
6791.5527 & 6.0166  & 0.6992 \\
6795.5095 & 7.2116  & 0.7634 \\
6800.4644 & 7.1856  & 0.6102 \\
6801.5000 & 5.8983  & 0.4060 \\
6802.5067 & 4.9370  & 0.5346 \\
6803.5294 & 6.3181  & 0.4875 \\
6806.5318 & 3.6377  & 1.1296 \\
6807.6147 & 4.3881  & 0.8387 \\
6809.4604 & 5.4800  & 0.4523 \\
6810.4839 & 6.5743  & 0.6446 \\
6811.4890 & 5.4027  & 0.3113 \\
6818.5059 & 5.9112  & 0.7276 \\
6819.5294 & 6.4559  & 0.7309 \\
6823.5395 & 5.9154  & 0.6874 \\
6824.5183 & 7.5002  & 0.4767 \\
6825.5069 & 5.7093  & 0.4808 \\
6829.4942 & 4.9555  & 0.7242 \\
6832.5015 & 5.5865  & 0.6765 \\
6833.5015 & 5.4094  & 0.5411 \\
6835.5347 & 5.4131  & 0.7196 \\
6836.4977 & 6.0169  & 0.4646 \\
6837.4984 & 5.8978  & 0.2831 \\
6838.5142 & 5.1887  & 0.4357 \\
6839.5007 & 5.1460  & 0.4750 \\
6840.4992 & 5.0470  & 0.6156 \\
6845.4618 & 5.2723  & 0.5693 \\
6847.4567 & 6.2708  & 0.5228 \\
6850.5099 & 5.0739  & 0.5690 \\
6855.4628 & 3.2880  & 0.8426 \\
6856.4542 & 5.3101  & 0.5070 \\
6857.4572 & 5.4737  & 0.6054 \\
6858.4560 & 5.6069  & 0.5388 \\
6859.4699 & 5.2633  & 0.6833 \\
6863.4696 & 6.1416  & 0.5253 \\
6864.4670 & 4.3139  & 0.4121 \\
6866.5170 & 2.9306  & 1.0689 \\
6867.4637 & 5.1125  & 0.6642 \\
6870.4592 & 3.9702  & 0.7236 \\
6871.4663 & 3.4837  & 0.7023 \\
6872.4634 & 4.8141  & 0.5248 \\
6873.4627 & 4.5870  & 0.5503 \\
6874.4542 & 6.1275  & 0.5268 \\
6876.4581 & 5.5267  & 1.2326 \\
6878.4816 & 7.1632  & 0.5748 \\
6879.4665 & 5.3268  & 0.4990 \\
6880.4665 & 5.7424  & 0.4804 \\
6881.4646 & 5.5631  & 0.6196 \\
6882.4610 & 6.3667  & 0.4998 \\
6883.4684 & 5.2834  & 0.5567 \\
6885.4659 & 3.5044  & 0.8973 \\
6886.4726 & 3.7979  & 0.7180 \\
6887.4599 & 5.4324  & 0.5923 \\
6941.9068 & 5.5135  & 0.6352 \\
6942.9073 & 6.0722  & 0.6552 \\
6943.9038 & 7.4447  & 0.5162 \\
6950.8754 & 7.1946  & 0.4945 \\
6951.8733 & 8.1639  & 0.5595 \\
6952.8645 & 8.1985  & 0.5905 \\
6953.8796 & 8.2420  & 0.6296 \\
6954.8312 & 9.3900  & 0.8081 \\
6955.8613 & 8.3603  & 0.5537 \\
6956.8738 & 9.1790  & 0.5338 \\
6957.8902 & 8.5548  & 0.3678 \\
6958.8639 & 7.9437  & 0.5608 \\
6959.8471 & 8.5141  & 0.5679 \\
6961.8481 & 9.3317  & 0.5193 \\
6964.8650 & 8.7717  & 0.5678 \\
6965.8623 & 8.9091  & 0.4967 \\
6968.8670 & 9.1311  & 0.5058 \\
6969.8629 & 8.4862  & 0.5019 \\
6972.8807 & 9.0138  & 0.5930 \\
6974.8472 & 9.0937  & 0.5025 \\
6977.8456 & 8.9742  & 0.5387 \\
6979.8393 & 7.3828  & 0.4877 \\
6980.8055 & 8.8080  & 0.6131 \\
6982.8125 & 9.3305  & 0.3660 \\
6984.8198 & 9.3482  & 0.5050 \\
6985.8277 & 9.0623  & 0.5774 \\
6986.8716 & 9.2466  & 0.5336 \\
6988.7943 & 8.7045  & 0.6594 \\
6990.7638 & 8.9411  & 0.6907 \\
6992.7814 & 9.3522  & 0.5176 \\
6994.8169 & 8.3602  & 0.4506 \\
6996.8516 & 7.8459  & 0.5787 \\
6998.8053 & 9.0233  & 0.5028 \\
7000.8371 & 9.0422  & 0.3881 \\
7007.8533 & 8.0849  & 0.5467 \\
7008.7387 & 9.1964  & 0.5310 \\
7009.7487 & 9.7889  & 0.4396 \\
7012.8169 & 8.1188  & 0.3972 \\
7013.8113 & 9.5207  & 0.4539 \\
7014.7384 & 9.6725  & 0.4726 \\
7015.7661 & 11.3078 & 0.5339 \\
7016.7963 & 10.3123 & 0.5559 \\
7018.7787 & 9.5129  & 0.4672 \\
7021.8358 & 9.3003  & 0.6022 \\
7022.7519 & 9.6649  & 0.7746 \\
7025.8280 & 9.2541  & 0.4911 \\
7026.7004 & 9.5918  & 0.4978 \\
7030.6911 & 10.5962 & 0.7018 \\
7032.8183 & 9.7752  & 0.4993 \\
7035.7243 & 9.6177  & 0.3830 \\
7037.8164 & 9.6682  & 0.6541 \\
7038.8521 & 9.6077  & 0.6162 \\
7039.8336 & 8.6459  & 0.4043 \\
7040.7121 & 9.6737  & 0.4904 \\
7043.8463 & 10.5003 & 0.5425 \\
7046.6847 & 9.5991  & 0.6591 \\
7047.6855 & 9.0888  & 0.7053 \\
7048.8103 & 9.1553  & 0.7067 \\
7050.7318 & 9.3499  & 0.7276 \\
7051.8465 & 9.5500  & 0.6863 \\
7059.7902 & 10.4699 & 0.4528 \\
7060.6626 & 10.3002 & 0.5673 \\
7061.8510 & 10.8566 & 0.6771 \\
7062.6967 & 9.5440  & 0.4022 \\
7063.7354 & 9.0539  & 0.5210 \\
7064.7755 & 10.3209 & 0.6899 \\
7068.7710 & 8.5098  & 0.4984 \\
7070.6468 & 11.1138 & 0.4044 \\
7074.8003 & 9.9561  & 0.5636 \\
7075.6999 & 10.5591 & 0.6446 \\
7078.8993 & 10.1916 & 0.4279 \\
7083.6618 & 11.6915 & 0.3053 \\
7088.6417 & 9.5766  & 0.5989 \\
7091.5928 & 11.7059 & 0.6554 \\
7092.7233 & 10.3560 & 0.4098 \\
7093.7108 & 10.5341 & 0.4324 \\
7095.8005 & 11.3241 & 0.6450 \\
7097.6962 & 9.0192  & 0.5465 \\
7104.5614 & 11.1620 & 0.4707 \\
7115.7489 & 11.3934 & 0.5647 \\
7118.5573 & 11.2683 & 0.4354 \\
7120.5391 & 10.2184 & 0.5184 \\
7124.4992 & 12.2814 & 0.5199 \\
7125.5235 & 9.8804  & 0.3943 \\
7127.6192 & 10.3576 & 0.3959 \\
7130.5443 & 10.3833 & 0.4921 \\
7132.5158 & 9.5261  & 0.6126 \\
7134.6264 & 10.3966 & 0.3892 \\
8116.8606 & 5.2999 & 0.4137 \\
8118.8165 & 5.4894 & 0.4418 \\
8121.8505 & 4.9621 & 0.4014 \\
8130.7972 & 5.2279 & 0.5254 \\
8132.7853 & 4.2357 & 0.5714 \\
8133.8294 & 4.8694 & 0.4552 \\
8135.7970 & 5.0718 & 0.4733 \\
8145.8177 & 5.7343 & 0.4658 \\
8146.8286 & 5.2924 & 0.4023 \\
8147.8066 & 5.9192 & 0.4301 \\
8148.7815 & 5.9017 & 0.4739 \\
8150.7582 & 5.3690 & 0.4648 \\
8158.6342 & 5.7879 & 0.4606 \\
8160.7501 & 5.8058 & 0.4439 \\
8162.7974 & 6.0305 & 0.4349 \\
8164.7352 & 5.5477 & 0.3781 \\
8172.7982 & 6.0642 & 0.4702 \\
8174.6811 & 5.8303 & 0.4847 \\
8176.6893 & 6.1190 & 0.4063 \\
8178.6189 & 6.1358 & 0.4895 \\
8186.7597 & 5.9310 & 0.4177 \\
8188.6939 & 5.6569 & 0.4698 \\
8190.6400 & 5.7723 & 0.4124 \\
8192.7439 & 5.8068 & 0.4421 \\
8200.6405 & 5.7340 & 0.4374 \\
8202.5807 & 5.7793 & 0.4124 \\
8204.6013 & 4.9772 & 0.4790 \\
8206.6028 & 4.9764 & 0.4171 \\
8214.5330 & 5.6293 & 0.5741 \\
8216.5715 & 5.2566 & 0.4795 \\
8218.5708 & 4.5351 & 0.5933 \\
8220.5664 & 5.1248 & 0.4728 \\
8228.6131 & 4.9383 & 0.6416 \\
8230.5611 & 4.8883 & 0.4610 \\
8232.5621 & 4.9927 & 0.4078 \\
8242.5156 & 4.8756 & 0.5288 \\
8244.5167 & 4.7232 & 0.6114 \\
8246.5215 & 4.1816 & 0.5869 \\
8248.6102 & 4.1594 & 0.6660 \\
8258.5731 & 4.6028 & 0.4507 \\
8260.4913 & 4.1079 & 0.6641 \\
8262.4987 & 4.7661 & 0.4661 \\
8270.5720 & 3.1171 & 1.0923 \\
8272.4622 & 4.4305 & 0.4789 \\
8274.5431 & 4.2690 & 0.6600 \\
8276.5334 & 3.7131 & 0.5313 \\
8284.5375 & 4.2254 & 0.6119 \\
8287.4957 & 4.8761 & 0.4310 \\
8289.5123 & 4.1456 & 0.6214 \\
8290.5663 & 4.9216 & 0.4633 \\
8298.5078 & 4.0474 & 0.5453 \\
8301.5267 & 4.8849 & 0.6908 \\
8303.4741 & 4.4906 & 0.5030 \\
8316.4768 & 4.4153 & 0.7053 \\
8329.4678 & 4.8609 & 0.6449 \\
8330.4688 & 4.7149 & 0.6765 \\
8452.8630 & 4.5234 & 0.4178 \\
8466.8571 & 4.4358 & 0.4701 \\
8480.8632 & 4.7315 & 0.5157 \\
8495.7624 & 3.6414 & 0.4933 \\
8513.6726 & 3.5280 & 0.5242 \\
8777.8660 & 5.3919 & 0.9074 \\
8782.8819 & 5.8229 & 0.6407 \\
8785.8744 & 4.7468 & 0.8103 \\
8800.8592 & 6.3867 & 0.4662 \\
8803.8497 & 6.1182 & 0.6688 \\
8805.8312 & 6.8611 & 0.6851 \\
8809.8619 & 6.5885 & 0.8066 \\
8812.8630 & 6.1588 & 0.5554 \\
8814.8574 & 6.4803 & 0.6234 \\
8817.8610 & 5.7883 & 0.5637 \\
8820.8104 & 6.1536 & 0.4876 \\
8822.8133 & 6.0344 & 0.5743 \\
8824.7950 & 6.0404 & 0.6142 \\
8826.8360 & 5.7774 & 0.5780 \\
8827.8608 & 6.2554 & 0.4869 \\
8830.8214 & 6.2087 & 0.8675 \\
8832.8593 & 6.9087 & 0.6375 \\
8834.8460 & 6.7220 & 0.6081 \\
8836.8174 & 5.8818 & 0.6183 \\
8852.8207 & 5.7680 & 0.7024 \\
8853.8375 & 6.5221 & 0.9110 \\
8854.8133 & 5.1232 & 0.9352 \\
8855.8436 & 5.9815 & 0.7943 \\
8856.8048 & 5.7298 & 0.6826 \\
8857.8663 & 5.2833 & 0.6781 \\
8860.8514 & 5.1155 & 0.5466 \\
8861.8533 & 6.2699 & 0.4861 \\
8862.8308 & 5.6169 & 0.5726 \\
8863.8699 & 5.6570 & 0.5895 \\
8864.8254 & 6.8291 & 0.6214 \\
8866.8648 & 5.3074 & 0.9421 \\
8867.8760 & 5.1028 & 1.2621 \\
8868.8816 & 5.4226 & 0.8160 \\
8871.8362 & 5.1718 & 0.7734 \\
8872.8191 & 5.1878 & 1.0159 \\
8875.8653 & 5.3220 & 0.6858 \\
8879.8167 & 5.3580 & 0.6493 \\
8880.8738 & 5.9770 & 0.7499 \\
8881.8639 & 5.4094 & 0.6616 \\
8882.8219 & 5.4810 & 0.6340 \\
8883.8449 & 5.9745 & 0.7162 \\
8884.7985 & 6.0315 & 0.6037 \\
8886.8221 & 5.4267 & 0.5306 \\
8887.8172 & 5.9484 & 0.4732 \\
8888.8627 & 5.9402 & 0.5647 \\
8889.8636 & 5.9201 & 0.5624 \\
8890.8608 & 5.2729 & 0.5544 \\
8891.8962 & 5.4412 & 0.5985 \\
8892.8507 & 4.8640 & 0.7056 \\
8893.8244 & 5.5802 & 0.5533 \\
8894.8883 & 5.0969 & 0.6055 \\
8895.7919 & 6.8546 & 0.7473 \\
8896.8512 & 5.1043 & 0.7421 \\
8897.7793 & 4.7744 & 0.5408 \\
8898.7409 & 4.9412 & 0.5014 \\
8899.7816 & 5.3626 & 0.5083 \\
8900.7151 & 5.4527 & 0.5108 \\
8901.7245 & 5.1093 & 0.5539 \\
8902.7389 & 5.4294 & 0.6548 \\
8903.7444 & 4.8602 & 0.5707 \\
8904.7448 & 5.0492 & 0.5340 \\
8905.7200 & 4.6773 & 0.5662 \\
8906.7154 & 5.2632 & 0.5417 \\
8907.7330 & 5.0922 & 0.5449 \\
8908.7334 & 5.0118 & 0.4672 \\
8909.7035 & 4.7002 & 0.5172 \\
8910.7072 & 6.0793 & 0.5556 \\
8911.6643 & 4.9627 & 0.5609 \\
8912.6790 & 4.9955 & 0.5705 \\
8914.7153 & 4.5219 & 0.6795 \\
8915.7889 & 5.7322 & 0.6033 \\
8916.7380 & 4.9506 & 0.6643 \\
8917.7593 & 6.3266 & 0.5666 \\
8918.6995 & 5.4386 & 0.5988 \\
8919.6849 & 5.6215 & 0.6395 \\
8920.7120 & 5.3495 & 0.5328 \\
8921.7181 & 5.4387 & 0.5980 \\
8922.6421 & 4.2238 & 0.8165 \\
8923.6323 & 5.0022 & 0.6467 \\
8924.6323 & 4.7297 & 0.5665 \\
8925.6034 & 5.2500 & 0.4730
\enddata
\end{deluxetable}

\startlongtable
\begin{deluxetable}{ccc}
\tablecaption{$-$ 168 \kms Measurements \label{table:168}}
\tablewidth{0pt}
\centering
\tablehead{
  \colhead{HJD}         &
  \colhead{$W_{\lambda}$}  &
  \colhead{$\sigma$}  \\  
  \colhead{-2,450,000}         &
  \colhead{}  &  
  \colhead{}         \\ 
  \colhead{(d)}  &
  \colhead{(m\AA)}  &
  \colhead{(m\AA)}
  }
\startdata
8116.8606 & 1.9270  & 2.0744 \\
8118.8165 & 2.7298  & 2.2157 \\
8121.8505 & 2.8522  & 2.0058 \\
8130.7972 & 3.1699  & 2.6286 \\
8132.7853 & 5.9941  & 2.8296 \\
8133.8294 & 3.0175  & 2.2726 \\
8135.7970 & 4.1528  & 2.3626 \\
8145.8177 & 2.9724  & 2.3386 \\
8146.8286 & 3.7882  & 2.0119 \\
8147.8066 & 3.0426  & 2.1619 \\
8148.7815 & 3.7487  & 2.3795 \\
8150.7582 & 5.8807  & 2.3191 \\
8158.6342 & 4.8814  & 2.3076 \\
8160.7501 & 4.5840  & 2.2253 \\
8162.7974 & 6.5519  & 2.1777 \\
8164.7352 & 6.0477  & 1.8886 \\
8172.7982 & 4.8312  & 2.3603 \\
8174.6811 & 4.9278  & 2.4292 \\
8176.6893 & 4.6321  & 2.0407 \\
8178.6189 & 6.9212  & 2.4517 \\
8186.7597 & 6.5458  & 2.0901 \\
8188.6939 & 5.0281  & 2.3515 \\
8190.6400 & 4.6226  & 2.0666 \\
8192.7439 & 6.2721  & 2.2111 \\
8200.6405 & 5.5918  & 2.1889 \\
8202.5807 & 5.0442  & 2.0657 \\
8204.6013 & 3.7935  & 2.3908 \\
8206.6028 & 4.1444  & 2.0806 \\
8214.5330 & 4.9041  & 2.8736 \\
8216.5715 & 4.5730  & 2.3948 \\
8218.5708 & 1.7513  & 2.9603 \\
8220.5664 & 3.9028  & 2.3617 \\
8228.6131 & 1.8869  & 3.2091 \\
8230.5611 & 1.9861  & 2.3046 \\
8232.5621 & 1.5371  & 2.0417 \\
8242.5156 & 3.1707  & 2.6394 \\
8244.5167 & 1.8863  & 3.0541 \\
8246.5215 & 1.5284  & 2.9227 \\
8248.6102 & 4.5160  & 3.3030 \\
8258.5731 & 3.2661  & 2.2456 \\
8260.4913 & 1.5134  & 3.3055 \\
8262.4987 & 5.1208  & 2.3189 \\
8270.5720 & 2.3722  & 5.3956 \\
8272.4622 & 4.1881  & 2.3804 \\
8274.5431 & 6.3511  & 3.2676 \\
8276.5334 & 2.2828  & 2.6348 \\
8284.5375 & 3.0182  & 3.0421 \\
8287.4957 & 3.3950  & 2.1505 \\
8289.5123 & 3.1830  & 3.0871 \\
8290.5663 & 3.3647  & 2.3125 \\
8298.5078 & 1.8562  & 2.7121 \\
8301.5267 & 1.5341  & 3.4561 \\
8303.4741 & 1.9703  & 2.5084 \\
8316.4768 & 0.6062  & 3.5221 \\
8329.4678 & 1.2385  & 3.2268 \\
8330.4688 & 1.4468  & 3.3811 \\
8452.8630 & 0.6586  & 2.0873 \\
8466.8571 & 0.8109  & 2.3471 \\
8480.8632 & 4.0911  & 2.5686 \\
8495.7624 & 3.6353  & 2.4410 \\
8513.6726 & -2.7691 & 2.6141 \\
8777.8660 & -5.9810 & 4.0175 \\
8782.8819 & -3.0090 & 2.8326 \\
8785.8744 & -2.9033 & 3.5581 \\
8800.8592 & -1.7399 & 2.0647 \\
8803.8497 & -1.1761 & 2.9545 \\
8805.8312 & 0.6667  & 3.0328 \\
8809.8619 & -0.0915 & 3.5684 \\
8812.8630 & 0.1391  & 2.4498 \\
8814.8574 & -2.3675 & 2.7647 \\
8817.8610 & 0.8139  & 2.4784 \\
8820.8104 & -1.0610 & 2.1543 \\
8822.8133 & -2.1444 & 2.5394 \\
8824.7950 & -1.8230 & 2.7145 \\
8826.8360 & -2.8950 & 2.5544 \\
8827.8608 & -3.0781 & 2.1587 \\
8830.8214 & -2.6295 & 3.8425 \\
8832.8593 & -4.6527 & 2.8437 \\
8834.8460 & -2.8550 & 2.7029 \\
8836.8174 & -2.4430 & 2.7322 \\
8852.8207 & -4.9671 & 3.1129 \\
8853.8375 & -6.8060 & 4.0663 \\
8854.8133 & -4.2678 & 4.1242 \\
8855.8436 & -4.4630 & 3.5223 \\
8856.8048 & -3.0746 & 3.0165 \\
8857.8663 & -5.2199 & 2.9973 \\
8860.8514 & -6.9145 & 2.4193 \\
8861.8533 & -5.2517 & 2.1616 \\
8862.8308 & -4.4471 & 2.5336 \\
8863.8699 & -4.7974 & 2.6102 \\
8864.8254 & -8.5891 & 2.7861 \\
8866.8648 & -3.8925 & 4.1570 \\
8867.8760 & -0.3216 & 5.5342 \\
8868.8816 & -3.8268 & 3.6030 \\
8871.8362 & -2.9384 & 3.4052 \\
8872.8191 & -0.2323 & 4.4564 \\
8875.8653 & -4.8696 & 3.0303 \\
8879.8167 & -2.7358 & 2.8614 \\
8880.8738 & -5.1024 & 3.3282 \\
8881.8639 & -5.5478 & 2.9277 \\
8882.8219 & -5.6743 & 2.8073 \\
8883.8449 & -0.6598 & 3.1591 \\
8884.7985 & -0.0741 & 2.6614 \\
8886.8221 & 3.9531  & 2.3174 \\
8887.8172 & 5.0786  & 2.0700 \\
8888.8627 & 6.3572  & 2.4656 \\
8889.8636 & 9.5681  & 2.4444 \\
8890.8608 & 10.0664 & 2.3980 \\
8891.8962 & 13.8973 & 2.5777 \\
8892.8507 & 13.7307 & 3.0286 \\
8893.8244 & 14.5997 & 2.3825 \\
8894.8883 & 18.3980 & 2.5852 \\
8895.7919 & 18.3334 & 3.2259 \\
8896.8512 & 17.4236 & 3.1734 \\
8897.7793 & 19.9830 & 2.2992 \\
8898.7409 & 21.6081 & 2.1290 \\
8899.7816 & 21.2090 & 2.1652 \\
8900.7151 & 20.9040 & 2.1779 \\
8901.7245 & 23.4560 & 2.3480 \\
8902.7389 & 21.8435 & 2.7880 \\
8903.7444 & 24.7157 & 2.4112 \\
8904.7448 & 21.7421 & 2.2686 \\
8905.7200 & 21.6379 & 2.4001 \\
8906.7154 & 23.8953 & 2.2970 \\
8907.7330 & 23.4400 & 2.3098 \\
8908.7334 & 21.2560 & 1.9855 \\
8909.7035 & 24.2127 & 2.1845 \\
8910.7072 & 22.5546 & 2.3726 \\
8911.6643 & 21.9787 & 2.3805 \\
8912.6790 & 21.3513 & 2.4242 \\
8914.7153 & 20.7693 & 2.8814 \\
8915.7889 & 22.1137 & 2.5725 \\
8916.7380 & 23.4706 & 2.8134 \\
8917.7593 & 22.4455 & 2.4234 \\
8918.6995 & 18.3622 & 2.5623 \\
8919.6849 & 19.5739 & 2.7347 \\
8920.7120 & 20.1201 & 2.2729 \\
8921.7181 & 18.0060 & 2.5603 \\
8922.6421 & 18.7528 & 3.4659 \\
8923.6323 & 18.1697 & 2.7605 \\
8924.6323 & 19.4647 & 2.4099 \\
8925.6034 & 17.4798 & 2.0244
\enddata
\end{deluxetable}

\end{document}